\documentclass[reprint,showpacs,showkeys,amsmath,amssymb,aps]{revtex4-1}	
\usepackage{natbib}
\usepackage{subfig}
\usepackage{graphicx}
\usepackage{xcolor}
\usepackage{tikz}
\usepackage{pgfplots}

\def\Xint#1{\mathchoice
{\XXint\displaystyle\textstyle{#1}}%
{\XXint\textstyle\scriptstyle{#1}}%
{\XXint\scriptstyle\scriptscriptstyle{#1}}%
{\XXint\scriptscriptstyle\scriptscriptstyle{#1}}%
\!\int}
\def\XXint#1#2#3{{\setbox0=\hbox{$#1{#2#3}{\int}$ }
\vcenter{\hbox{$#2#3$ }}\kern-.6\wd0}}

\def\dashint{\Xint-}

\begin{document}

\preprint{Drafted for {\em Physical Review E}, \today}
\title{Saffman--Taylor fingers with kinetic undercooling}

\author{Bennett P.~J.~Gardiner}
\author{Scott W. McCue}
\email{scott.mccue@qut.edu.au}
\affiliation{Mathematical Sciences, Queensland University of Technology, Brisbane QLD 4001, Australia}

\author{Michael C. Dallaston}
\affiliation{Mathematical Institute, University of Oxford, Andrew Wiles Building,
Radcliffe Observatory Quarter,
Oxford OX2 6GG, UK}

\author{Timothy J. Moroney}
\affiliation{Mathematical Sciences, Queensland University of Technology, Brisbane, QLD 4001, Australia}

\begin{abstract}

The mathematical model of a steadily propagating Saffman--Taylor finger in a Hele-Shaw channel has applications to two-dimensional interacting streamer discharges which are aligned in a periodic array.  In the streamer context, the relevant regularisation on the interface is not provided by surface tension, but instead has been postulated to involve a mechanism equivalent to kinetic undercooling, which acts to penalise high velocities and prevent blow-up of the unregularised solution.  Previous asymptotic results for the Hele-Shaw finger problem with kinetic undercooling suggest that for a given value of the kinetic undercooling parameter, there is a discrete set of possible finger shapes, each analytic at the nose and occupying a different fraction of the channel width. In the limit in which the kinetic undercooling parameter vanishes, the fraction for each family approaches $1/2$, suggesting that this `selection' of $1/2$ by kinetic undercooling is qualitatively similar to the well-known analogue with surface tension.  We treat the numerical problem of computing these Saffman--Taylor fingers with kinetic undercooling, which turns out to be more subtle than the analogue with surface tension, since kinetic undercooling permits finger shapes which are corner-free but not analytic.  We provide numerical evidence for the selection mechanism by setting up a problem with both kinetic undercooling and surface tension, and numerically taking the limit that the surface tension vanishes.

\end{abstract}


\maketitle


\section{Introduction}

Inspired by the seminal work of Saffman and Taylor \cite{ST}, an enormous amount of research has been undertaken on aspects of the problem of a steadily moving finger of inviscid fluid in a Hele-Shaw cell of channel geometry (for an overview of Hele-Shaw flows, with a thorough discussion on flows in the channel geometry, see \cite{Tanveer00,Vasilev09,AmyTsai12}).  In a typical experiment \cite{ST, Tabeling87}, air is injected from the left end of a horizontal channel, which is otherwise filled with viscous fluid.  The air-fluid interface is unstable, as the less viscous fluid is displacing the more viscous fluid (the Saffman--Taylor instability).  As the interface evolves from left to right, a fingering pattern develops, which ultimately results in a single finger of air propagating steadily along the Hele-Shaw cell and occupying a fraction $\lambda\in (1/2,1)$ of the channel width. As the finger speed increases (via higher injection rates), the ratio $\lambda$ is observed to decrease towards roughly $\lambda=1/2$ \cite{ST}.

The most common approach to understanding the structure of the corresponding mathematical model is to study the role of a dimensionless surface tension parameter $\sigma$, which decreases as the finger speed increases \cite{MS}.  There are exact solutions for the special case $\sigma=0$ \cite{ST,MS,H}, but these all take the ratio $\lambda$ as an input parameter, and so do not describe the observed experimental behaviour.  The `selection' of $\lambda=1/2$ as the physically appropriate solution in the limit $\sigma\rightarrow 0$ is a difficult problem in exponential asymptotics \cite{Combescot86,Combescot88,Dorsey87,Hong86,Shraiman86,Tanveer87,C}.  The relevant analysis predicts that, for a given $\sigma$, there is a discrete set of solutions with $1/2<\lambda<1$. As $\sigma\rightarrow 0$, the value of $\lambda$ for each solution branch approaches the special value $\lambda=1/2$. Numerical solutions support these conclusions \cite{MS,VDB}.

In the present study, we are concerned with the effect that kinetic undercooling has on the Hele-Shaw problem in a channel geometry.  The appropriate dimensionless model for a steadily propagating finger is \cite{CK}
\begin{subequations}
\begin{align}
\qquad \nabla^2\phi & =0 \quad & \mbox{in} & \quad  \Omega_{\infty},&   \label{eq:Laplacetw} \\
\qquad \frac{\partial\phi}{\partial n} & =0 \quad & \mbox{on}& \quad \partial\Omega_{\infty},&   \label{eq:kinematictw} \\
\qquad \phi  &=cv_n-\frac{x}{1-\lambda} \quad & \mbox{on} &\quad \partial\Omega_{\infty},&  \label{eq:dynamictw}  \\
\qquad \frac{\partial\phi}{\partial y} & =0 \quad & \mbox{on}& \quad y=\pm 1,&   \label{eq:Neumanntw} \\
\qquad \phi & \sim -\frac{x}{1-\lambda}  \quad & \mbox{as}& \quad x  \rightarrow -\infty, \nonumber \\
& & &  \lambda<|y|<1, &  \label{eq:farupstreamtw}\\
\qquad  \phi & \sim -x \quad & \mbox{as} & \quad  x\rightarrow +\infty, &   \nonumber \\
& & & \!\!\!\!\! -1<y<1. & \label{eq:fardownstreamtw}
\end{align}
\label{eq:goveqns}
\end{subequations}
Here $\phi$ is the velocity potential in the frame of reference of the finger, $\partial/\partial n$ denotes a directional derivative normal to the interface $\partial\Omega_{\infty}$, $v_n$ is the normal velocity of the interface, and $c$ is the kinetic undercooling parameter. The unregularised version (zero kinetic undercooling, $c=0$) of Eqs.~(\ref{eq:goveqns}) has Eq.~(\ref{eq:dynamictw}) replaced by
\begin{equation} \label{eq:dynamictw2}
\phi=-\frac{x}{1-\lambda}
\quad\mbox{on}\quad \partial\Omega_{\infty},
\end{equation}
which also applies for the zero surface tension case mentioned above.

Kinetic undercooling-type conditions arise in a variety of applications. In the Hele-Shaw context, the kinetic undercooling term arises from the curvature in the transverse direction (perpendicular to the parallel walls of the Hele-Shaw cell), and its dependence on the interface velocity. This effect was included by Romero \cite{Romero}, who modelled the contact angle as a linear function of the velocity, leading to a boundary condition such as Eq.~(\ref{eq:dynamictw}). An alternative interpretation is to consider the existence of a wetting layer of the receding fluid that remains on the plates of the Hele-Shaw cell. Park and Homsy \cite{Park1984} derived a power-law relationship between the thickness of this layer and the capillary number. This relationship leads to a power-law dependence on velocity, with the term $cv_n$ in Eq.~(\ref{eq:dynamictw}) replaced by $cv_n^{\gamma}$, where $\gamma = 2/3$ is the exponent derived in \cite{Park1984}.  Such a term may be referred to as representing \emph{nonlinear kinetic undercooling}.  The theoretical short-time existence of solutions to Hele-Shaw flow with this regularisation was established by Pleshchinskii and Reissig \cite{Plesh2002}. Recently, the stability of an expanding circular bubble with both surface tension and nonlinear kinetic undercooling has been considered, in both linear \cite{Marty2008,Dias2013}, and weakly nonlinear \cite{Anjos2013,Anjos2014} regimes. In this paper, however, we consider linear kinetic undercooling ($\gamma=1$) only.

In the context of melting or freezing, Stefan-type formulations may include a Gibbs-Thomson law with kinetic undercooling \cite{King2006, Chen1997, Back2014,Back2014b}, with much attention given to instabilities and pattern formation at the interface of a growing dendrite \cite{Misbah91,Chen1997,Gibou03}; in that case, in the limit of vanishingly small specific heat, the governing equations reduce to those for Hele-Shaw flow.  Thus the unstable Hele-Shaw model describes the manner in which a supercooled liquid freezes, with Eqs.~(\ref{eq:goveqns}) above relevant for a single dendrite propagating with constant velocity in a channel.  Kinetic undercooling conditions also apply on interfaces in very similar moving boundary problems describing mass transfer situations, such as the diffusion of solvent through glassy polymers \cite{McCue2011, Mitchell2014}.

Of particular interest here, the model (\ref{eq:goveqns}) has applications to streamers, which is a topic that has received much attention in the physics literature in recent times (see the review \cite{Ebert11}). Streamers are finger-shaped electrical discharges which occur during the early stages of electric breakdown in sparks or lightning, for example.  They are caused by subjecting a weakly ionized gas to a strong electric field, leading to an ionization reaction via collisions of highly energetic electrons with neutral molecules.  The streamers themselves are characterized by a thin charge layer and associated ionization front that forms the finger shape.

A minimal model for streamer discharges consists of a coupled system of reaction diffusion equations for the electron and ion density.  A further equation relates the Laplacian of the electrostatic potential $\phi$ to these densities. For negative streamers, these equations can be approximated by a moving boundary problem by assuming the ionization layer is a sharp interface that separates the strongly ionized streamers from the weakly ionized gas ahead of front.  The result is Laplace's equation for the electrostatic potential outside the interface.
For the case in which there is a periodic array of two-dimensional streamers with equal spacing, all propagating in the $x$-direction with a constant electric field ${\bf E}=-x {\bf i}$ in the far field as $x\rightarrow\infty$, one can impose Neumann conditions to isolate a single streamer \cite{Luque08,Ebert11,Kyuregyan14}.  Under this periodic geometry, if the electric field or periodic spacing is sufficiently small (strong interaction between neighbouring streamers), the streamers evolve from their initial conditions to a travelling wave profiles, so that they propagate uniformly.  The approximate model is then given by Eqs.~(\ref{eq:goveqns}).

In the context of streamers, the boundary condition (\ref{eq:dynamictw2}) has been used instead of Eq.~(\ref{eq:dynamictw}) (see \cite{Meulenbroek04}, for example).  The former is appropriate if the streamer is assumed to be ideally conducting ($\phi=0$ in the streamer) and the electric potential is assumed to be continuous across the interface.  Indeed, the condition (\ref{eq:dynamictw2}) was used by Luque et al~\cite{Luque08} in their study of periodic streamers (see also Ref.~\cite{Kyuregyan14}).  However, as is known from the Hele-Shaw literature, the unregularised time-dependent model is ill-posed, with a dense subset of all initial conditions leading to finite time blow-up that is characterised by infinitely sharp cusps on the interface \cite{H}.  Such behaviour is not physical (in either the Hele-Shaw or streamer context).  The regularising term (\ref{eq:dynamictw}) is postulated by Ebert and coworkers \cite{Meulenbroek05,Ebert07,Brau08} for streamers, and used, for example, to model perturbed translating circles \cite{Tanveer2009,Kao2010}. A further relevant discussion is contained in Ref.~\cite{Ebert11}.  Here the kinetic undercooling parameter $c$ is proposed to account for the thickness of the ionization front.  In the present paper, we shall employ the language of Hele-Shaw flows, but keep in mind the application of streamers, discussing the relevance of the analysis and results in Sec.~\ref{discuss}.

The Saffman--Taylor problem with kinetic undercooling, described by Eqs.~(\ref{eq:goveqns}), has received modest attention compared to the surface tension analogue mentioned above.  The selection problem was treated by Chapman and King \cite{CK}, who used exponential asymptotics to show that discrete families of analytic fingers exist, with the finger width for each family tending to $1/2$ in the limit that the kinetic undercooling parameter $c$ vanishes.  These authors showed that $\lambda\sim 1/2+\alpha c^{2/3}$ as $c\rightarrow 0$ for each branch, but did not compute the constant $\alpha$. More recently, a numerical study by Dallaston and McCue~\cite{DM} showed that, for a given kinetic undercooling parameter $c$, a continuous family of corner-free finger solutions exist with widths $\lambda \in [\lambda_\mathrm{min},1)$. Further, it was found that the minimum width $\lambda_\mathrm{min} \rightarrow 0$ as $c\rightarrow 0$. While this continuous spectrum of solutions appears to be at odds with the analysis of Chapman and King~\cite{CK}, the two studies need not contradict each other since the numerical scheme in Ref.~\cite{DM} is not designed to distinguish between solutions with analytic fingers and those with fingers that are also corner-free but may not be analytic (that is, for all the numerical solutions found in Ref.~\cite{DM}, the first derivative exists at the nose, but the higher order derivatives may not exist there).

In this paper we aim to reconcile these results by constructing numerical solutions to Eqs.~(\ref{eq:goveqns}) that have analytic fingers. The rigorous results of Tanveer and Xie~\cite{TX,XT} suggest that solutions to the Hele-Shaw problem with sufficiently small values of the surface tension coefficient must have interfaces that are analytic.  With this in mind, our strategy is add surface tension to the model (\ref{eq:goveqns}), so that Eq.~(\ref{eq:dynamictw}) is replaced by
\begin{equation}
\phi=\sigma\kappa+cv_n-\frac{x}{1-\lambda}
\quad\mbox{on}\quad \partial\Omega_{\infty},\label{dynboth}
\end{equation}
where $\sigma$ is the surface tension coefficient and $\gamma$ is the curvature of the interface (see Ref.~\cite{Dallaston2013} for an in-depth study of Hele-Shaw flows with surface tension and kinetic undercooling). Our hypothesis is that the work of Tanveer and Xie carries over to Eqs. (\ref{eq:Laplacetw})--(\ref{eq:kinematictw}), (\ref{eq:Neumanntw})--(\ref{eq:fardownstreamtw}), (\ref{dynboth}) so that solutions to the problem with kinetic undercooling and surface tension must be analytic at the nose.  Thus with kinetic undercooling fixed at some value $c>0$, by taking the limit $\sigma\rightarrow 0$, we select the analytic solutions studied in Chapman and King~\cite{CK}.  Using this strategy, we are able to produce a plot of finger widths $\lambda$ versus kinetic undercooling $c$ for the first two branches, thus filling in the gap left by Chapman and King~\cite{CK} and Dallaston and McCue~\cite{DM}.  Our results have implications for the problem of periodic streamers studied by Luque et al~\cite{Luque08}.

Our numerical scheme is based on a boundary integral formulation, as outlined in Sec.~\ref{BIF}. Sec.~\ref{Numerics} summarises our main results, while Sec.~\ref{discuss} includes a discussion.

\section{Boundary integral formulation}\label{BIF}

For the formulation of the problem, we follow the work of McLean and Saffman \cite{MS} and Chapman and King \cite{C,CK}. Since $\phi$ is a harmonic function, we define an analytic complex potential $w(z) = \phi(x,y)+ \mathrm{i}\psi(x,y)$, where $\psi$ is a stream function and $z = x+\mathrm{i}y$. The conformal transformation $z \mapsto w$ maps the fluid region onto an infinite strip of unit width in the potential plane. A second conformal map, $w \mapsto \chi = \xi +\mathrm{i}\eta = \mathrm{e}^{-\pi w}$ maps this strip onto the upper half $\chi$-plane. The interface is mapped onto the unit interval on the real line, $0<\xi<1$, with the upper wall mapped onto $-\infty < \xi < 0$ and the center line $y=0$ mapped onto $1<\xi<\infty$.

The complex velocity can be written
\begin{equation}
\frac{\partial  w}{\partial z} = \mathbf{\hat{\text{$q$}}} \mathrm{e}^{-\mathrm{i}\hat{\theta}}
\end{equation}
where $\mathbf{\hat{\text{$q$}}}$ is the velocity tangential to streamlines, and $\hat{\theta}$ is the angle the tangent to the streamlines make with the $x$-axis.  The logarithm of this velocity, $\log{\mathbf{\hat{\text{$q$}}}} - \mathrm{i}\hat{\theta}$, is analytic in the upper half $\chi$-plane, and its real and imaginary parts can be related by a property of Hilbert transforms called the Kramers-Kronig relations, such that
\begin{equation} \label{qint}
\log{\mathbf{\hat{\text{$q$}}}} = -\frac{1}{\pi} \dashint_{0}^{1} \frac{\hat{\theta}(\xi') - \pi}{\xi'-\xi} \ \mathrm{d}  \xi', \qquad 0 < \xi < 1,
\end{equation}
since $\hat{\theta} = \pi$ everywhere on the real line except the unit interval. Note that the integral is of Cauchy principal value type.

Relating the quantities $\mathbf{\hat{\text{$q$}}}(\xi)$ and $ \hat{\theta}(\xi)$ to the curvature of the interface (Ref.~\cite{MS}) allows us to rewrite the dynamic condition (\ref{dynboth}) as the differential equation
\begin{align}
(1-\lambda)\mathbf{\hat{\text{$q$}}} = & (1-\lambda)\pi^2 \sigma q \xi  \frac{\mathrm{d}}{\mathrm{d} \xi} \left( \mathbf{\hat{\text{$q$}}} \xi \frac{\mathrm{d} \hat{\theta}}{\mathrm{d} \xi} \right)
\nonumber
\\
& +c\pi \mathbf{\hat{\text{$q$}}} \xi \,\cos \hat{\theta}\frac{\mathrm{d} \hat{\theta}}{\mathrm{d}\xi} - \cos \hat{\theta}, \qquad 0 < \xi <1.
\label{de}
\end{align}
We now have Eqs.~\eqref{qint} and \eqref{de} relating $\mathbf{\hat{\text{$q$}}}$ and $\hat{\theta}$, with the associated boundary conditions
\begin{equation}
\hat{\theta}(0) = \pi, \quad \mathbf{\hat{\text{$q$}}}(0) = \frac{1}{1-\lambda}, \quad
\hat{\theta}(1) = \frac{\pi}{2}, \quad \mathbf{\hat{\text{$q$}}}(1)  = 0,
\label{bcs1}
\end{equation}
which correspond to uniform flow at the tail $(\xi=0)$ and a stagnation point at the nose $(\xi=1)$.

Given values of the physical parameters $\sigma$ and $c$, we seek to solve Eqs~(\ref{qint})-(\ref{bcs1}), and then compute the finger width $\lambda$ via
\begin{equation} \label{lamcalc}
\log{\left(1-\lambda\right)} = \frac{1}{\pi} \int_{0}^{1} \frac{\hat{\theta}(\xi') - \pi}{\xi'} \ \mathrm{d} \xi',
\end{equation}
which comes from setting $\xi=0$ into Eqn~(\ref{qint}).

We now introduce another variable substitution that simplifies the equations and removes the explicit dependence on $\lambda$. We let $\theta(\xi) = \hat{\theta}(\xi)-\pi$, $q(\xi) = (1-\lambda)\mathbf{\hat{\text{$q$}}(\xi)}$ and introduce new parameters
\begin{equation}
\gamma = \frac{\sigma \pi^2}{1-\lambda}, \quad \epsilon = \frac{c\pi}{2(1-\lambda)}
\end{equation}
where $\gamma$ is a scaled surface tension parameter \cite{MS,C} and $\epsilon$ is a scaled kinetic undercooling parameter \cite{CK}. Then with some manipulation, the governing equations become
\begin{align}
q &=\gamma q \xi \frac{\mathrm{d} }{\mathrm{d} \xi} \left\{q \xi \frac{\mathrm{d} \theta}{\mathrm{d} \xi}\right\} + 2\epsilon q \xi \cos \theta \frac{\mathrm{d} \theta}{\mathrm{d} \xi}  + \cos \theta,
\label{ST_ODE} \\
\log q &= - \frac{\xi}{\pi} \dashint_{0}^{1} \frac{\theta(\xi')}{\xi'(\xi'-\xi)} \ \mathrm{d} \xi',   \label{integral}
\end{align}
both of which hold on $0 < \xi <1$, together with boundary conditions
\begin{equation}
\theta(0) = 0, \quad q(0) =1, \quad \theta(1) =-\frac{\pi}{2}, \quad q(1)=0.
\label{BCs}
\end{equation}
Given a solution for $\theta(\xi)$, we can calculate the width of the finger using Eq.~(\ref{lamcalc}), which is now
\begin{equation}
\log (1-\lambda) = \frac{1}{\pi} \int_{0}^{1} \frac{\theta(\xi')}{\xi'} \ \mathrm{d} \xi',
\end{equation}
and calculate the shape of the interface from
\begin{equation}
x(\xi)+\mathrm{i}y(\xi) =- \frac{1-\lambda}{\pi} \int_{\xi}^{1} \frac{\exp\left(\mathrm{i} \theta(\xi')\right)}{q(\xi')\xi'} \ \mathrm{d} \xi'.  \label{coords}
\end{equation}

Using McLean and Saffman's exact solutions for the unregularised problem \cite{MS},
\begin{equation}
q=\left(\frac{1-\xi}{1+a \xi}\right)^{1/2},
\quad
\theta = \cos^{-1} q,
\label{eq:exactsoln}
\end{equation}
where $a = (2\lambda -1)/(1-\lambda)^2$ is arbitrary, and the formulae for the physical coordinates implicit in Eq.~(\ref{coords}), we can recover the analytic formula for the shape of the finger given by Saffman and Taylor, namely that $x(\xi) = ((1-\lambda)/\pi) \log \xi$, $y(\xi) = (2\lambda/\pi) \cos^{-1}\sqrt{\xi}$.  Combining the two results gives
\begin{equation}
x = \frac{2(1-\lambda)}{\pi} \log \left(\cos \left(\frac{\pi y}{2\lambda} \right) \right),
\end{equation}
which is often referred to in the literature as the ZST solution, being equivalent to the expression derived first by Zhuravlev \cite{Zhuravlev1956} and then by Saffman and Taylor \cite{ST} (see \cite{Tanveer00} for an alternative derivation).

\section{Numerical results}\label{Numerics}
We solve our system of integro-differential equations (\ref{ST_ODE})-(\ref{BCs}) by applying the numerical scheme outlined in the Appendix.  The approach involves dividing the domain $0<\xi<1$ into $N+1$ unevenly spaced grid points and solving a system of $N-1$ equations for the unknown function $\theta$ at each of the $N-1$ interior points using a Newton solver. The other quantities of interest can be computed subsequently.

A consequence of discretising the integral in Eq.~(\ref{integral}) is that the $N-1$ equations depend on the unknown function $\theta$ at {\em all} of the grid points, which leads to a fully dense Jacobian ${\bf J}$ in the Newton scheme. In order to proceed with a large number of grid points, we have employed a Jacobian-free Newton-Krylov method which does not require the formation of the full Jacobian; instead, a sparse approximation is all that is required for preconditioning of the Krylov subspace linear solver, as described in the Appendix.

Typically, for a fixed surface tension parameter $\gamma>0$ and kinetic undercooling parameter $\epsilon \ge 0$, the scheme converged to a solution that corresponds to a particular finger shape with a single finger width $\lambda$.  The initial guess used for Newton's method was either the exact solution (\ref{eq:exactsoln}) for $\gamma=0$, $\epsilon=0$, or an already converged solution with similar parameter values.  For moderate to large values of $\gamma$, $N=3000$ grid points were used, while for small values of $\gamma$ we used a larger number of grid points, up to a maximum of $N=5000$.

\begin{figure}%
  \centering
  \subfloat[]{\includegraphics[width=0.4\textwidth]{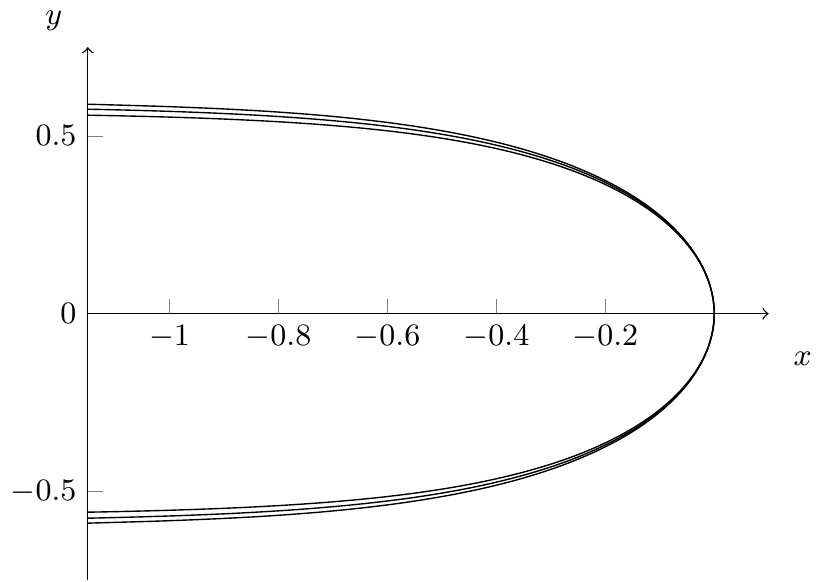} \label{fingers}}%

\hspace{-0.5cm}
  \subfloat[]{ \includegraphics[width=0.38\textwidth]{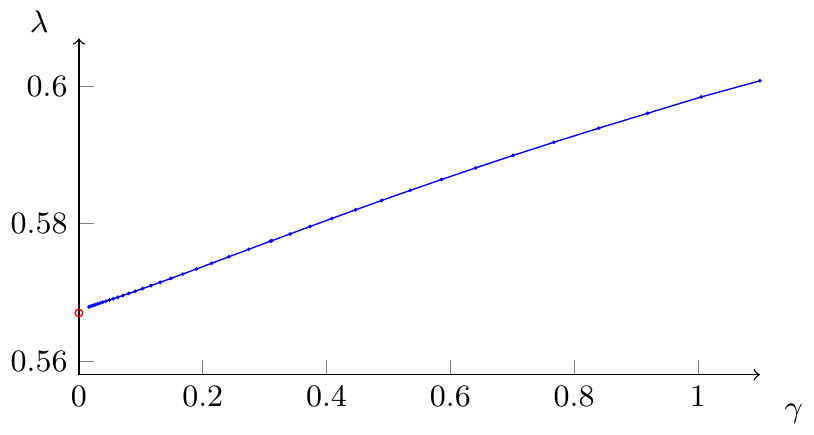} \label{primary}}
  \caption{(Color online) Numerical results for kinetic undercooling $\epsilon = 0.1$. (a) The shape of the fingers for the surface tension values $\gamma=0.03,0.5,1$ (from innermost to outermost curve). (b) The dependence of the finger width $\lambda$ on surface tension $\gamma$. Numerically computed data points are indicated by the solid (blue) circles. The open (red) circle is the estimated finger width for this family at $\gamma = 0$.}%
  \label{fig:test}%
\end{figure}

Some representative finger shapes are presented in Fig.~\subref{fingers}. Here we have fixed the kinetic undercooling parameter to be $\epsilon=0.1$ and provided results for three different surface tension values, $\gamma=0.03, 0.5$ and $1$.  We observe that the fingers are qualitatively the same in each case, and that the finger width $\lambda$ is greater than $1/2$ and decreases as the surface tension $\gamma$ decreases.

Each of these three solutions correspond to a single data point on the curve in Fig.~\subref{primary}, which shows the dependence of the finger width $\lambda$ on the surface tension $\gamma$ for $\epsilon=0.1$.  This figure clearly demonstrates the trend that as surface tension decreases, the finger width decreases.  For values of surface tension below roughly $\gamma\approx 0.015$, we were unable to compute sufficiently well converged solutions (using up to $N=5000$ grid points).  The reason for this breakdown in the numerical scheme is related to the singular nature of the limit $\gamma\rightarrow 0$, which is illustrated by the highest derivative in Eq.~(\ref{ST_ODE}) being multiplied by $\gamma$.  We discuss this issue further below.

Recall that our hypothesis is that all these fingers are analytic curves, since non-zero surface tension does not allow non-analytic solutions.  On the other hand, for $\gamma=0$ (zero surface tension), Dallaston and McCue \cite{DM} show that there is a continuous family of corner free solutions for $\lambda>\lambda_{\mathrm{min}}(\epsilon)$, where for $\epsilon=0.1$ the minimum value is roughly $\lambda_{\mathrm{min}}\approx 0.44$.  To select a single solution in this family (with $\epsilon = 0.1$ and $\gamma=0$) that has an analytic finger, we propose to consider the branch of solutions for $\epsilon=0.1$ and $\gamma>0$ and take the limit $\gamma\rightarrow 0^+$.

Since it is difficult to calculate solutions for extremely small values of $\gamma$, we use an extrapolation approach to obtain an estimate for the finger width at $\gamma=0$. One option to achieve this is to fit a polynomial to the last few data points and extract the value of this polynomial at $\gamma=0$. However, we have the result in the case of zero kinetic undercooling that $\lambda \sim \frac{1}{2} + \beta \gamma^{2/3}$ as $\gamma \rightarrow 0^+$, thus it seems reasonable to suggest that the same scaling holds in the case of finite kinetic undercooling. As such, we use the relation
\begin{equation}
\lambda \sim \alpha  + \beta \gamma^{2/3}, \label{extrap_eqn}
\end{equation}
and fit a small number of the final few points to this equation. The value obtained for $\alpha$ is the predicted finger width for $\gamma=0$, the intercept on the vertical axis in Fig.~\subref{primary}.

In addition to the branch of solutions shown in Fig.~\subref{primary}, we have found evidence of additional solution branches.  This is precisely the same behaviour as known to occur for the case without kinetic undercooling ($\epsilon=0$) \cite{C,VDB,Romero}. Romero \cite{Romero} and Vanden-Broeck \cite{VDB} demonstrated the existence of multiple solution branches for a given value of $\gamma$ numerically, and Chapman \cite{C} and others proved the existence of an infinite number of branches using exponential asymptotics. Kessler and Levine \cite{Kessler1,Kessler2} suggested that only the lower branch is stable while the other, higher branches are unstable \cite{Tanveer87,Tanveer00}.

\begin{figure}%
  \centering

  \subfloat[$\epsilon=0$]{\includegraphics[width=0.45\textwidth]{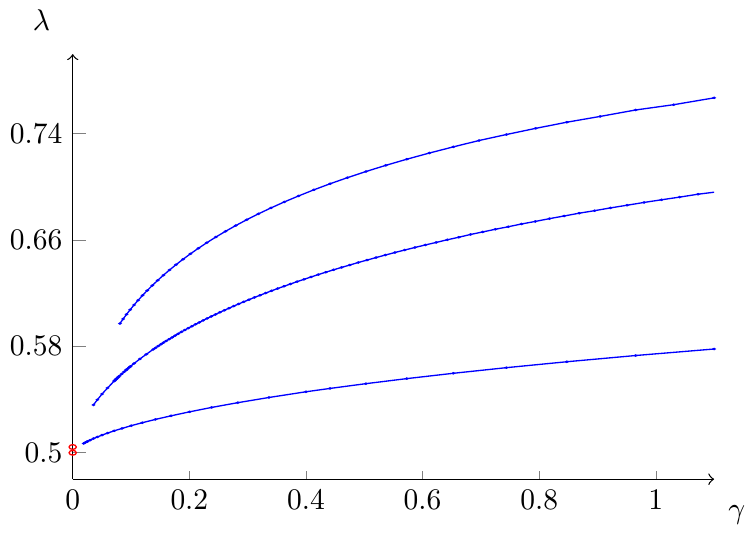} \label{e0}}%

  \subfloat[$\epsilon=0.1$]{ \includegraphics[width=0.45\textwidth]{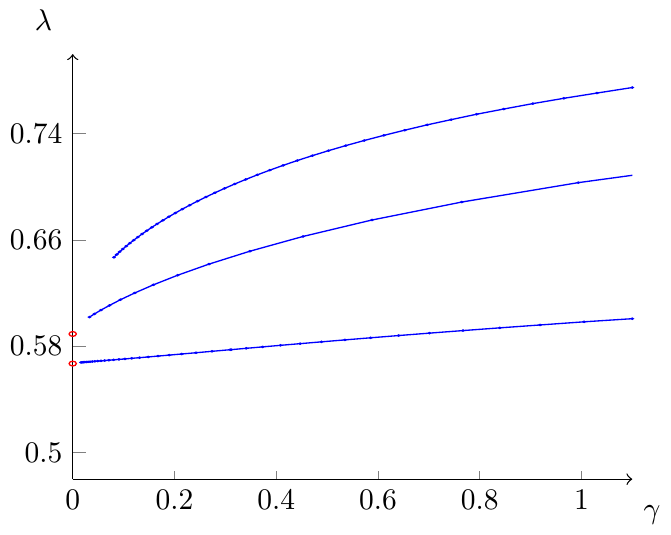} \label{e0point1}}

  \subfloat[$\epsilon=0.2$]{ \includegraphics[width=0.45\textwidth]{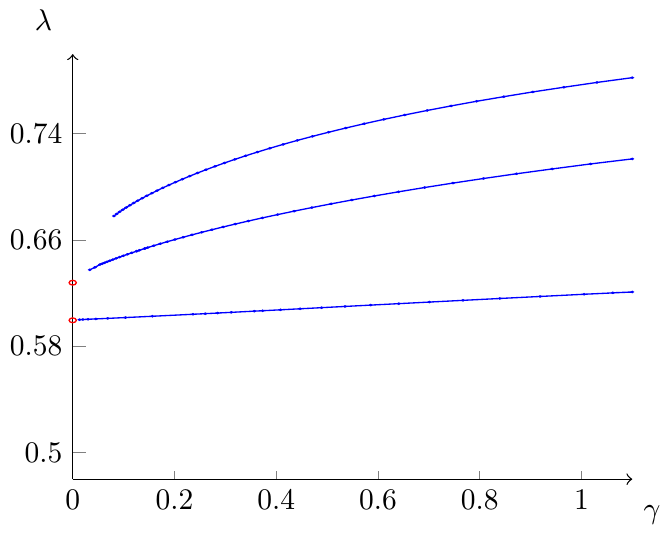} \label{e0point2}}

\caption{(Color online) Dependence of finger width $\lambda$ on surface tension $\gamma$ for fixed values of kinetic undercooling.  In each case, three solutions branches are shown.  The open (red) circles represent an extrapolated value for $\gamma=0$.}
\label{STtozero}
\end{figure}

Thus for this particular example $\epsilon=0.1$, we postulate there are a countably infinite number of solutions branches, each more difficult to compute than the previous. We show three such curves in Fig.~\subref{e0point1}.  Each follows the trend of decreasing $\lambda$ as $\gamma$ decreases. It is difficult to compute $\lambda$ values for small values of $\gamma$, but again, we are able to extrapolate to estimate the analytic solution for $\gamma=0$ on a second branch.  For the third branch, the lowest $\gamma$ value at which the numerical scheme converged was too large to give an accurate extrapolation estimate.

Also shown in Figs.~\ref{STtozero}(a) and \ref{STtozero}(c) are three solutions branches for $\epsilon = 0$ and $\epsilon=0.2$, respectively. Of course, the $\epsilon = 0$ case is the original surface tension problem \cite{MS,VDB,C}. The extrapolation technique was used on two branches for each of these values to obtain an estimate for an analytic solution $\gamma=0$. In principal we could construct a similar figure for any fixed value of kinetic undercooling, $\epsilon$.

\begin{figure}%
  \centering

  \subfloat[$\epsilon=0$]{\includegraphics[width=0.45\textwidth]{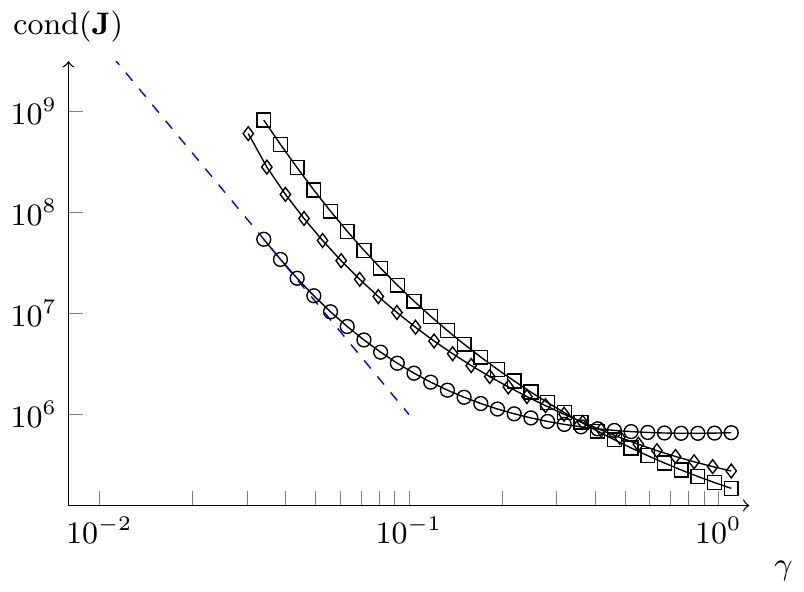} \label{Je0}}%

  \subfloat[$\epsilon=0.1$]{ \includegraphics[width=0.45\textwidth]{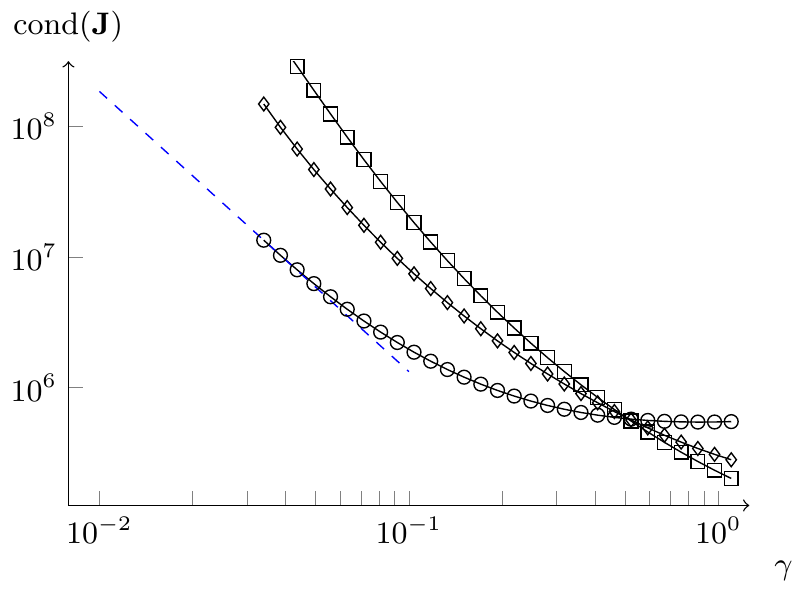} \label{Je0point1}}

  \subfloat[$\epsilon=0.2$]{ \includegraphics[width=0.45\textwidth]{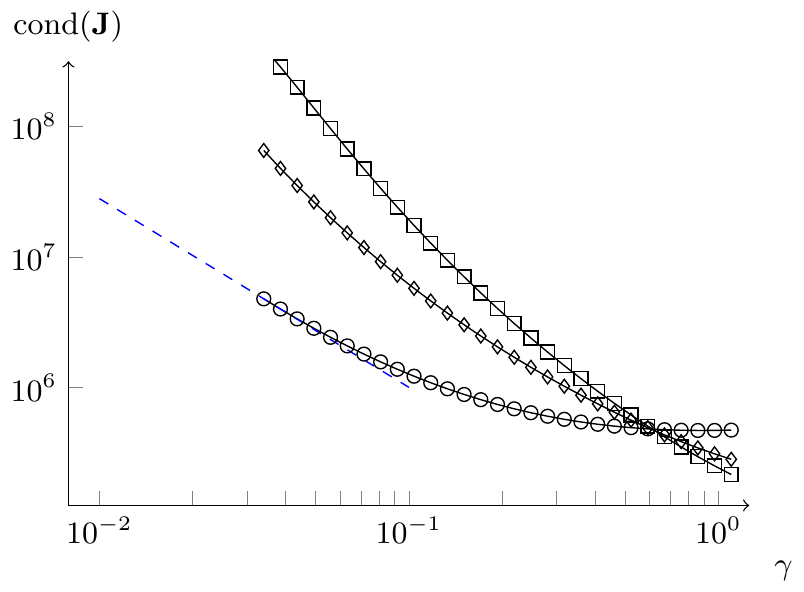} \label{Je0point2}}

\caption{(Color online) The condition number of the Jacobian for solutions computed with $N=1000$ nodes, plotted against $\gamma$ for the first branch (circles), second branch (diamonds) and third branch (squares), for fixed values of $\epsilon$. The (blue) dashed line is a rough linear fit through the data on the first branch for small $\epsilon$.}
\label{Jcond}
\end{figure}

To provide further insight into the singular nature of the limit $\gamma\rightarrow 0$, we have presented in Fig.~\ref{Jcond} plots of the 2-norm condition number, $\operatorname{cond}{({\bf J})}$, of the Jacobian versus the surface tension $\gamma$ for the cases presented in Fig.~\ref{STtozero}.  On this log-log scale, the data appears to be linear as $\gamma \rightarrow 0$, which implies that $\operatorname{cond}{({\bf J})}\sim \mbox{const}\,\gamma^{-p}$, where $p$ is a positive constant depending only on the kinetic undercooling parameter $\epsilon$ and the particular branch of solution.  These observations support the claim that the problem is singular in the limit $\gamma\rightarrow 0$ for $\epsilon\geq 0$, which is consistent with our hypothesis that the findings of Tanveer and Xie \cite{TX,XT} do not extend to the case $\gamma=0$, $\epsilon>0$. The singular nature of the problem also helps to explain the numerical findings \cite{DM} of a continuous solution space for $\gamma=0$, $\epsilon>0$.

\begin{figure}%
  \centering

  \subfloat[]{\includegraphics[width=0.45\textwidth]{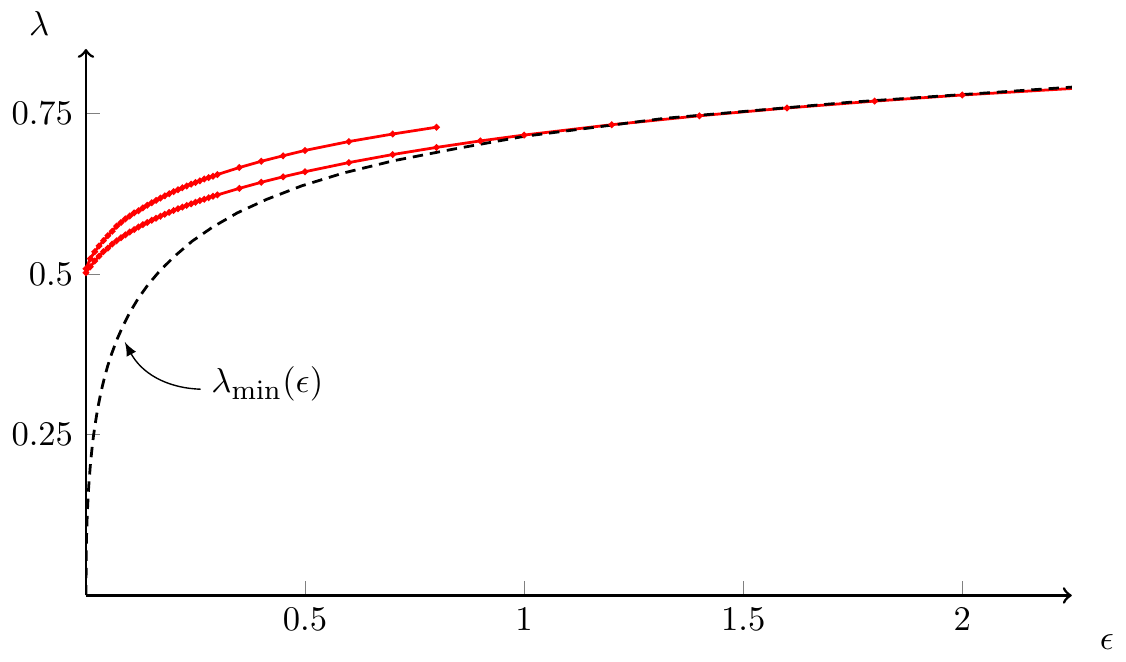} \label{fig:mina}}%

  \subfloat[]{ \includegraphics[width=0.45\textwidth]{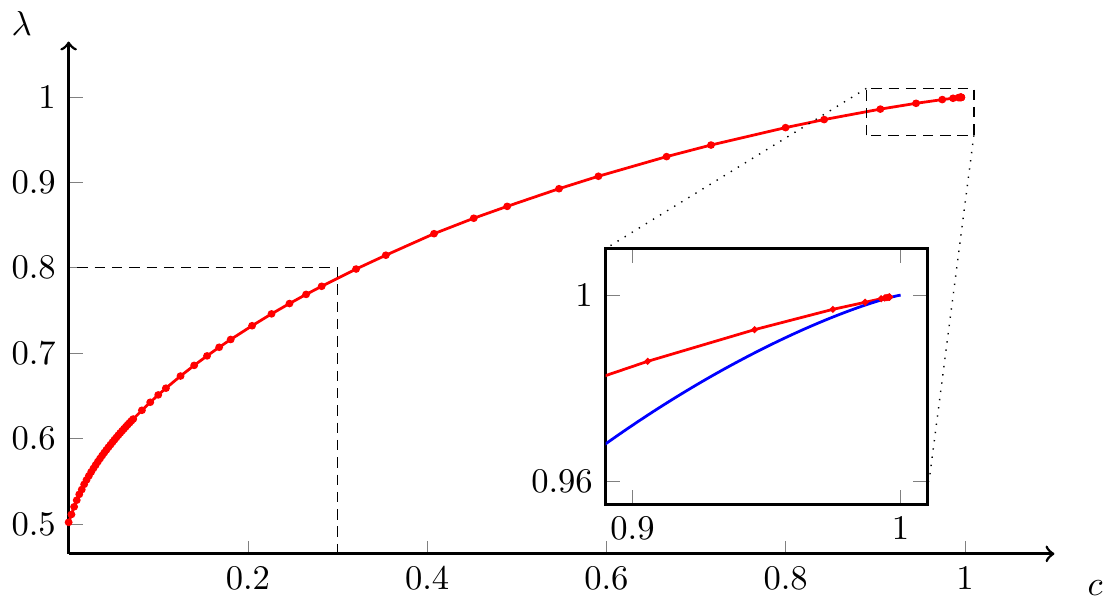} \label{fig:minb}}

\caption{(Color online) This figure constitutes our main result.  (a)  The selection of two distinct branches shown as the solid (red) circles. Note the distinction between these and the continuous solution space found by Dallaston and McCue \cite{DM}, bounded below by $\lambda_{\mbox{\footnotesize{min}}}(\epsilon)$, the dashed (black) curve. The primary branch seems to asymptote to the curve $\lambda_{\mbox{\footnotesize{min}}}(\varepsilon)$ as $\epsilon \rightarrow \infty$. \\ (b) The entire primary branch from part (a). Recall that as $\epsilon \rightarrow \infty$, $c \rightarrow 1^-$ and $\lambda \rightarrow 1^-$. The portion of the branch shown in part (a) is boxed in the lower left corner for reference. The inset shows a comparison with Eq.~(\ref{asymp_relation}) provided by Chapman \& King \cite{CK}, shown as the smooth (blue) curve.}
\label{fig:min}
\end{figure}

By extrapolating our numerical results for the lower two solution branches for many different values of $\epsilon$, we have constructed the data provided in Fig.~\subref{fig:mina}. These are our estimates of the finger widths associated with the analytic solutions to Eqs.~(\ref{eq:goveqns}), also analysed in Chapman \& King \cite{CK} using asymptotic techniques.  It is noteworthy that our solution branches in Fig.~\subref{fig:mina} also appear to approach $\lambda=1/2$ in the limit $\epsilon \rightarrow 0^+$, which agrees with Chapman \& King.

Also included in Fig.~\subref{fig:mina} as a dashed curve is the lower bound of all solutions, including non-analytic fingers, as found by Dallaston and McCue \cite{DM}. We see that as $\epsilon$ increases, this lower bound appears to asymptote to the lower solution branch for analytic fingers.

In Fig.~\subref{fig:minb} we include more details of the primary branch, showing the dependence of $\lambda$ against $0 \le c < 1$. Recall that as $\epsilon \rightarrow \infty$, $c \rightarrow 1^-$ and $\lambda \rightarrow 1^-$. Since our method is most useful for investigating the primary few branches in the ($\lambda, \epsilon$) solution space, we shall not attempt to match our curves to the solution curves in \cite{CK}, which are only valid near $\epsilon =0$ in the limit that $\epsilon N \rightarrow 0 $, where $N$ is the branch number, that is when $\lambda - 1/2 \sim \mathcal{O}(1)$. This implies that their results are only valid for the higher order branches. Unfortunately, it is therefore infeasible to use our proposed method to investigate these solution branches.

\begin{figure}%
  \centering

{\includegraphics[width=0.45\textwidth]{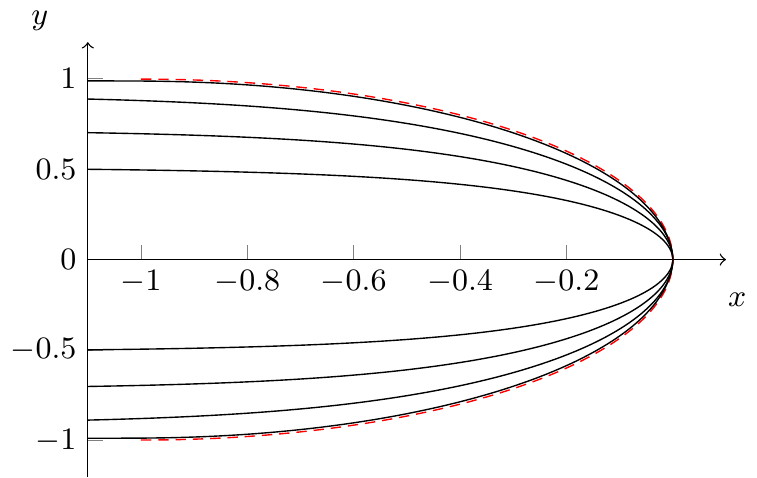}}

\caption{(Color online) The shape of the finger for solutions on the primary branch with $\gamma = 0.03$ and $\epsilon = 0 , 1, 10, 3500$ (from innermost to outermost curve). For comparison, the dashed (red) curve shows a semi-circle of unit radius.}
\label{KUfingers}
\end{figure}

While our main focus is selection as $\epsilon \rightarrow 0^+$, there are interesting results in the limit that the kinetic undercooling parameter $\epsilon \rightarrow \infty$, or equivalently, as $c \rightarrow 1^{-}$, which we can use to test our approach.  Chapman and King determined in the appendix of \cite{CK} that $1-\lambda \ll 1-c$ and that the asymptotic behaviour of the first branch is given by
\begin{equation}\label{asymp_relation}
c \sim 1-(1-\lambda) \log \left( 1/(1-\lambda) \right), \quad \mbox{as} \quad \lambda \rightarrow 1^-.
\end{equation}
See the inset in Fig.~\subref{fig:minb} for a comparison of the numerical results with this asymptotic relation. The shape of the finger in this limit is given by Chapman \& King \cite{CK} as being circular at the nose. See Fig.~\ref{KUfingers} for a comparison between this asymptotic solution and solution profiles for small surface tension and varying values of kinetic undercooling.

We end this section by mentioning the results published in Ref.~\cite{CTAC}.  The model~(\ref{eq:goveqns}) was treated in Ref.~\cite{CTAC} and numerical results were presented with discrete branches of solutions.  However, as discussed in detail by Dallaston and McCue~\cite{DM}, these discrete branches were due to very small numerical errors, which were corrected in Ref~\cite{DM}.

\section{Discussion and Conclusions} \label{discuss}
We have computed multiple discrete analytic solution branches for the Saffman--Taylor finger with kinetic undercooling, corresponding to those predicted asymptotically by Chapman and King \cite{CK}.  The greatest numerical challenge is to distinguish analytic solutions from non-analytic ones, given the inability of a finite difference scheme to capture high derivatives. Here we achieved this goal by extending the numerical scheme of \cite{MS,VDB} to include both surface tension and kinetic undercooling, and then extrapolating to find the limit as surface tension goes to zero for fixed kinetic undercooling values.  Our numerical results agree with asymptotic results in selecting a finger width of $1/2$ as $\epsilon \rightarrow 0^+$, as well as in producing a semi-circular interface as the finger width tends to the channel width.

The inclusion of surface tension ensures that the numerical solutions we compute represent analytic fingers; the theory of Tanveer and Xie \cite{TX} for the pure surface tension problem ensures that any solution that is $C^2$ (which finite differences can certainly distinguish) is also analytic, and it reasonable to assume this carries over when both surface tension and kinetic undercooling are present. The inability of the numerical method to distinguish discrete solution branches in the absence of surface tension (as observed in Ref.~\cite{DM}) suggests that the results of Tanveer and Xie do not apply when $\gamma=0$; that is, there do exist $C^2$ but nonanalytic travelling finger solutions for the pure kinetic undercooling problem.

We have not considered the numerical computation of the time-dependent version of Eqs.~(\ref{eq:goveqns}).  Analytic travelling finger solutions are only relevant if analyticity is preserved in evolving from an initial condition.  While this is unlikely to occur for sufficiently large kinetic undercooling (Dallaston and McCue \cite{DM} have numerical and asymptotic evidence of corner formation for $c > 1$) it may be possible if kinetic undercooling is small enough ($c<1$). Extrapolating a time dependent solution with zero surface tension and nonzero kinetic undercooling from one with nonzero surface tension and nonzero kinetic undercooling may introduce further complications given the structural instability of the time-dependent problem in the zero surface tension limit \cite{Tanveer00,Siegel96}. Any numerical scheme would have to be very precise, but also avoid the node-crowding effect typical of numerical conformal mapping methods.

We close with remarks about the relevance of our results for the study of streamer discharges.  For this application, it has been proposed that Hele-Shaw type models can be used to approximate the dynamic evolution of streamers, with a kinetic undercooling term used as a form of regularisation, where the kinetic undercooling parameter is a measure of the actual thickness of the ionization front \cite{Ebert11,Meulenbroek05,Ebert07,Brau08,Tanveer2009,Kao2010}.  Recall that Luque et al.~\cite{Luque08} considered a periodic array of strongly interacting streamers and showed that, after some transient period, they propagate uniformly.  By isolating a single translating streamer, they treated the Hele-Shaw problem (\ref{eq:goveqns}) as an approximate model, except that they used Eq.~(\ref{eq:dynamictw}) instead of (\ref{eq:dynamictw2}).  That is, they considered the unregularised version of the classical Saffman--Taylor finger problem \cite{ST}.  Here we have treated Eqs.~(\ref{eq:goveqns}) with nonzero kinetic undercooling, and presented results that support the hypothesis that the width of each streamer finger for vanishingly small kinetic undercooling (vanishingly small thickness of the ionization front) is one half the period of the array of periodic streamers~\cite{CK}.  This conclusion explains why the exact solution to the unregularised problem with the free parameter $\lambda$ set to $1/2$ agrees with time-dependent solutions to the full streamer problem, at least near the tip of the streamer~\cite{Luque08}.

As our study suggests, the use of a kinetic undercooling type regularisation for evolving streamers is not without complications.  While the Hele-Shaw model without regularisation is ill-posed, and therefore not appropriate for streamer discharges (or any application, for that matter), the time-dependent version of Eqs.~(\ref{eq:goveqns}) is still difficult to handle numerically.  For example, the time-dependent version of (\ref{eq:goveqns}) is highly unstable; linear stability shows \textit{all} modes of perturbation (of a flat interface) are unstable \cite{CK,H2}.  Further, it would presumably require a particularly sophisticated numerical scheme to distinguish between time-dependent solutions with analytic fingers and those that are non-analytic but corner-free.  As such, it seems that a better dynamic model for streamers may involve kinetic undercooling plus another regularisation effect that comes from the full streamer model.  This additional effect may then act like surface tension does in the Hele-Shaw context described here, allowing for selection of physically appropriate solutions to the streamer problem of interest.

\section*{Acknowledgements}
SWM acknowledges the support of the Australian Research Council via the Discovery Project DP140100933. MD acknowledges support in part by Award No KUK-C1-013-04, made by King Abdullah University of Science and Technology (KAUST). The authors acknowledge helpful discussions with John King and Jon Chapman.

\appendix*
\section{Numerical scheme}

We seek to solve the integro-differential set of equations (\ref{ST_ODE})-(\ref{integral}) and associated boundary conditions (\ref{BCs}) numerically, in a manner similar to McLean and Saffman \cite{MS} (see also Refs~\cite{VDB,DM,CTAC}).  In order to deal effectively with the integral in Eqn.~(\ref{integral}), we note that both $\theta$ and $q$ are non-differential functions of $\xi$ at the endpoints, with square root type singularities at $\xi=0$ (the tail) and $\xi =1 $ (the nose). The variable transformation
\begin{equation}
\xi^{\tau} = 1 - \zeta^{2},
\label{transform}
\end{equation}
is used to ensure that both variables have at least two derivatives at the end points, and $0<\tau<1/2$ is the real root of the transcendental equation
\begin{equation}
\gamma \tau^2 +2\epsilon\tau = \cot {\pi \tau},
\label{trans_eqn}
\end{equation}
which is obtained from considerations regarding the differentiability of $\theta$ at both endpoints \cite{MS}.

The integral in Eq.~(\ref{integral}) is a Cauchy Principal value integral; we can add and subtract  the singular part to give
\begin{align}
\log q &=  \frac{1}{\pi} \int_{0}^{1} \frac{\theta(\xi')}{\xi'} \ \mathrm{d} \xi' - \frac{1}{\pi}\int_{0}^{1} \frac{\theta(\xi')-\theta(\xi)} {\xi'-\xi} \ \mathrm{d} \xi' \nonumber \\
&\hspace{0.4cm}-\frac{\theta(\xi)}{\pi} \ln \left( \frac{1-\xi}{\xi} \right). \label{q_int_num}
\end{align}
Using Eq.~(\ref{transform}), the first integral in Eq.~(\ref{q_int_num}) becomes
$$
\frac{2}{\tau} \int_{0}^{1} \frac{\zeta' \theta(\zeta')}{1-\zeta'^{2}} \ \mathrm{d} \zeta'.
$$
Since now $\theta= 0$ at $\zeta=1$, the integrand has a removable singularity there, and can be replaced by $-(1/2)\left.\mathrm{d} \theta/\mathrm{d} \zeta \right|_{\zeta = 1}$.  Again using Eq.~(\ref{transform}), the second integral in Eq.~(\ref{q_int_num}) becomes
\begin{equation} \nonumber
\frac{2}{\tau} \int_{0}^{1}  \frac{\zeta'}{(1-\zeta'^{2})^{1-1/\tau}} \frac{\theta(\zeta') - \theta(\zeta)}{(1-\zeta'^{2})^{1/\tau} - (1-\zeta^{2})^{1/\tau}} \ \mathrm{d} \zeta',
\end{equation}
which has a removable singularity at $\zeta'=\zeta$. L'Hopital's rule is again used to replace the integrand at $\zeta'=\zeta$ with $-(\tau/2)\mathrm{d} \theta/\mathrm{d} \zeta$.

Now turning to the numerical scheme itself, we discretise the unit interval $\zeta\in [0,1]$ using $N+1$ nodes, $0\le \zeta_n \le 1$ where $n=0,1,2,\ldots,N$, and look to solve for the vector of unknowns $\mathbf{u} = [\theta_1, \theta_2, \ldots, \theta_{N-1}]^{\small{T}}$.  Given an initial guess $\mathbf{u}_0$ for the values of $\theta_n$, or an updated vector $\mathbf{u}_k$, we can calculate the values $[q_1, q_2, \ldots, q_{N-1}]^{\small{T}}$ using Eq.~(\ref{integral}) (rewritten in terms of $\zeta$), then substitute both $\theta$ and $q$ into Eq.~(\ref{ST_ODE}) using third order mixed finite difference formulas to approximate the derivatives.  Thus we have a system of $N-1$ nonlinear algebraic equations for the $N-1$ unknowns in $\mathbf{u}$, which we solve using a Jacobian-free Newton-Krylov method \cite{Knoll04}, implemented by the SUNDIALS package KINSOL~\cite{Hindmarsh}.  Once converged, the solution can be used to recalculate $q$, $\lambda$ and the physical coordinates $x$ and $y$.

\begin{figure}
  \centering

{\includegraphics[width=0.49\textwidth]{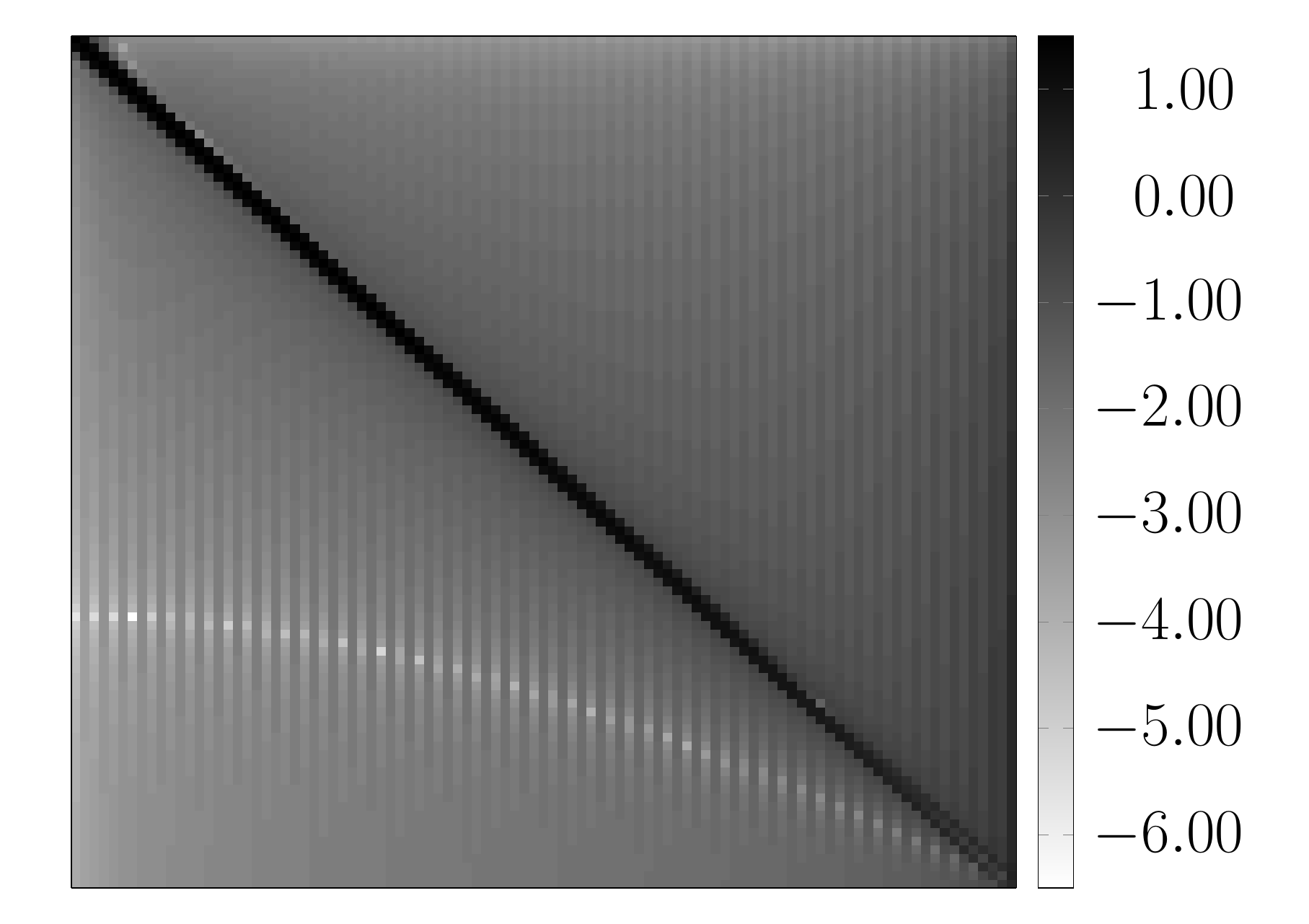}}

\caption{The structure of a typical Jacobian, for $N=100$. The plot shows $\log_{10}|\mathbf{J}|$, for $\gamma = 0.02, \epsilon =0.2$, on the first solution branch.}
\label{Jacobian}
\end{figure}

Due to the global nature of the integral equation~(\ref{integral}), the Jacobian matrix ${\bf J}$ of the nonlinear system is fully dense.  The Jacobian-free Newton-Krylov method avoids the need to form this dense matrix, leading to considerable efficiency gains.  It does so by using a preconditioned Krylov subspace solver at the linear level, which requires only an approximation of the true Jacobian for preconditioning purposes.  To efficiently construct this approximation, we observe that the largest entries in the Jacobian matrix are contained within a narrow band around the main diagonal -- a consequence of the finite difference approximation of the derivatives in Eq.~(\ref{ST_ODE}); other relatively large values are located in the rightmost columns.  An example of this striking pattern is provided in Fig.~\ref{Jacobian}, where we see the magnitude of the entries in ${\bf J}$ decay with distance from the main diagonal.  To construct the preconditioner, we retain only the entries within the narrow band and a relatively small number of the rightmost columns, yielding a sparse approximation that is efficient to form and factorise.  By varying the bandwidth, the trade-off between the cost of factorisation and the effectiveness of the preconditioner can be controlled.  This approach is analogous to that applied recently by Pethiyagoda et al.~\cite{Ravi1,Ravi2}, who also solved a coupled system of two integro-differential equations, derived using a boundary integral method.  Similar tactics for constructing sparse preconditioners from dense Jacobians have been implemented for other non-local systems (see Ref.~\cite{moroney13}).

\end{document}